\documentclass[preprint,english,showpacs,11pt,floatfix]{revtex4}
\usepackage{makeidx}
\usepackage{babel,amsmath,amssymb,dcolumn}
\usepackage{hyperref}
\usepackage{amsfonts}
\usepackage{amssymb}
\usepackage{graphicx}
\usepackage{latexsym}
\usepackage{dcolumn}

\newcommand{\be}{\begin{equation}}
\newcommand{\ee}{\end{equation}}

\newcommand{\bear}{\begin{eqnarray}}
\newcommand{\eear}{\end{eqnarray}}
\def\bearst{\begin{eqnarray*}}
\def\eearst{\end{eqnarray*}}
\begin{document}

\title{Integrable models: from dynamical solutions to string theory}
\author{Elcio Abdalla}
\email{eabdalla@fma.if.usp.br}
\affiliation{Instituto de F\'\i sica, Universidade de S\~ao Paulo, C.P.66.318, CEP
05315-970, Sao Paulo, Brasil}
\author{Antonio Lima-Santos}
\email{dals@df.ufscar.br}
\affiliation{Universidade Federal de S\~ao Carlos, Departamento de F\'{\i}sica \\
Caixa Postal 676, 13569-905, S\~{a}o Carlos-SP, Brasil}

\begin{abstract}
We review the status of integrable models from the point of view of their
dynamics and integrability conditions. Some integrable models are discussed
in detail. We comment on the use it is made of them in string theory. We
also discuss the SO(6) symmetric Hamiltonian with SO(6) boundary.

This work is especially prepared for the seventieth anniversaries of Andr%
\'{e} Swieca (in memoriam) and Roland K\"{o}berle.
\end{abstract}

\maketitle

\section{Introduction}

As a natural extension of Quantum Mechanics, Relativistic Quantum Field
Theory (QFT) has demonstrated its predictive power in the calculation of
processes in Quantum Electrodynamics. There are, however, conceptual and
technical difficulties, since the local products of quantum fields, which
are operator-valued distributions, are ill defined. This problem can only be
resolved via the techniques of renormalization.

The general non-perturbative properties of quantum field theory were first
extracted from a perturbative setup by the so-called LSZ formalism. Next,
dispersion relations were found and were used to obtain non-perturbative
information. These developments were followed by the axiomatic approach,
known as constructive QFT. An important consequence of this approach is the
CPT theorem connecting spin and statistics.

However, dynamical calculations in QFT were, in the sixties, restricted to
perturbation theory. Therefore, calculations involving strong interactions
were unreliable. Information information about the bound state spectrum were
very poor and could only be obtained within crude approximate schemes. Thus,
QFT fell into stagnation for many years. These difficulties provided a
motivation for the S-matrix theory. But its predictive power turned out to
be very small, since it was entirely based on kinematical principles,
analyticity and the bootstrap idea. An underlying dynamical framework was
lacking. Nevertheless, analyticity in the complex angular momentum plane led
to the important concept of duality. An explicit realization of these
concepts by the Veneziano formula led to a new parallel development in the
sixties, the dual models. However, the predictions of the dual models for
high-energy scattering processes were incorrect.

On the other hand, QFT explained very successfully the weak interactions.
Moreover, symmetry principles had proven powerful in predicting the masses
of strongly interacting particles without the recourse to dynamical
calculations. These facts led to a revival of QFT in the late sixties. In
the seventies, much effort has been spent on non-perturbative aspects.
Quantum Chromodynamics (QCD) was proposed as the fundamental theory of the
strong interactions as a result of the successful perturbative explanation
of high energy scattering as well as the success of the quark model.
Nevertheless, reliable non-perturbative calculations were still lacking in
four dimensions and were only available for specific models in
two-dimensional space-time\cite{aar}. It was understood that the short
distance singularities of quantum field theory play a key role in the
dynamical structure of the theory. The experimental results on lepton-proton
scattering at large momentum transfer, required that a realistic theory of
the strong interactions be asymptotically free.

The recourse to soluble or almost soluble models as a laboratory was a must
for a dynamical understanding of QFT. The first soluble model was that
describing a two dimensional massless fermion with a current-current
interaction formulated by Thirring in 1958 \cite{thirring} as an example of
a completely soluble quantum field theoretic model obeying the general
principles of a $QFT$\cite{klaiber}. Subsequently, Schwinger\cite{schwinger}
obtained an exact solution of Quantum Electrodynamics in 1+1 dimensions, $%
QED_2$. A number of interesting properties, such as the nontrivial vacuum
structure of this model, were understood only later\cite{loswie} when it was
found that there is a long range Coulomb force for the charge sectors of the
theory. This long range force was interpreted as being responsible for the
confinement of quarks\cite{cakosuss}. The problem of confinement and the
related phenomenon of screening of charge quantum numbers in two dimensions
have been studied by several authors\cite{screening1,screening2}, and have
served as a basis for understanding important concepts in QFT. The
surprisingly rich structure of two-dimensional quantum electrodynamics was
found to describe several important features of the non-abelian gauge
theories, which were under investigation in the seventies.

Several results of increasing importance followed. Two-dimensional
classically integrable models were studied in great detail. Such models are
characterized by the existence of an infinite number of conservation laws.
If these conservation laws survive quantization, the corresponding
S-matrices can be computed exactly\cite{36}. Some of the results concerning
classical integrability have also been generalized to higher dimensions\cite%
{41} and used to understand QCD\cite{fadeevkor,others}.

Describing two dimensional fermions in terms of bosons (bosonization) can
lead to non-perturbative information. The building blocks of the procedure
are the exponentials of the free bosonic fields. One obtains a fermion
number which is connected to the infrared behaviour of the massless scalar
fields. One thus obtains a superselection rule\cite{49} and the charged
sectors appear in a natural way.

A particularly important class of two-dimensional integrable non-linear
sigma models are those with a geometrical origin\cite{eichenherr}, which
share several properties with four dimensional Yang-Mills theories\cite%
{eichenherr,dadda}. Upon quantization they exhibit dynamical mass generation
and contain a long range force\cite{dadda} for simple gauge groups\cite%
{abdforgo}. Such a long range force can be screened by dynamical fermions 
\cite{75}. These properties make them appealing as toy models for the strong
interactions\cite{57}. They are also very interesting mathematical objects,
particularly important in the framework of string theory.

Furthermore, the study of these models has led to new developments in the
study of quantum field theories in higher dimensions. High-energy scattering
amplitudes involving fields with definite helicity or at high energy, in
four-dimensional Quantum Chromodynamics, have a rather simple description,
related to integrable models \cite{fadeevkor,others}. In the former case,
the scattering amplitudes are related to solutions of self-dual Yang-Mills
equation, while in the latter case the interaction of external particles is
described by the two-dimensional Heisenberg Hamiltonian of spin systems.

\section{Exact S Matrices and Yang Baxter equations}

The most general invariance group of a non trivial field theory in $d>2$
dimensions is the product of the Poincar\'{e} group and an internal symmetry%
\cite{[288;8]} times supersymmetry \cite{[289;8]}.

The basic idea of the proof is that an infinite number of higher
conservation laws implies that the momenta involved in the scattering
process are individually conserved, so that the process merely consists in
an exchange of quantum numbers. This would imply that the S-matrix does not
depend analytically on the scattering momenta; in particular, the
two-particle S-matrix would not depend analytically on the scattering angle.

In two-dimensional space-time the situation is different. The scattering
angle can only be zero or $\pi $ and the clash with analyticity no longer
exists. The constraints due to the conservation laws on the scattering
process are very strong. The conservation of an infinite number of local
charges implies conservation of the energy, momentum and their powers: the
higher conserved charges are higher-rank tensors $Q_{\mu _1 \cdots \mu _l}$,
transforming according to higher representations of the Lorentz group,
commuting with one another and with the momentum \cite{[36;8]}.

The action of $Q_{\mu _{1}\cdots \mu _{n}}$ on asymptotic states is severely
restricted by Lorentz invariance. On a one-particle state, we have 
\begin{equation}
Q^{\mu _{1}\cdots \mu _{l}}|P\rangle =P^{\mu _{1}}\cdots P^{\mu
_{l}}|P\rangle \quad .
\end{equation}%
Conservation of the higher charges thus imply 
\begin{equation}
\sum_{i=1}^{n}P_{i}^{\mu _{1}}\cdots P_{i}^{\mu
_{l}}=\sum_{i=1}^{m}P_{i}^{^{\prime }\mu _{1}}\cdots P_{i}^{^{\prime }\mu
_{l}}\quad ,  \label{8.3}
\end{equation}%
provided the corresponding scattering amplitudes do not vanish. Hence there
exists an infinite number of conservation laws 
\index{Conservation laws!and S-matrix} that must be obeyed by the external
momenta. Equations such as (\ref{8.3}) can only be satisfied if $n=m$, 
\textit{i.e.}, if there is no particle production, and the individual
momenta are conserved. Thus, after a suitable rearrangement, $%
P_{i}=P_{i}^{\prime }$, and the scattering only consists of time delays and
exchange of quantum numbers \cite{[36;8],[190;8]}. (We have ignored terms
such as $g^{\mu \nu }P^{\rho }|P^{a}\rangle $ since they are not essential.
Notice also that since the mass operator commutes with the charge $Q^{\mu
\nu \rho }$ there can be degeneracy.) 

\subsection{Factorizable S-matrix}

Absence of particle production implies that the S-matrix is of the
factorizable type, that is, the scattering S-matrix is given by the product
of all possible two-particle scattering amplitudes\cite{[190;8]}.
Furthermore, the two-particle processes are severely constrained by the
so-called \textit{factorization relations}.

In order to see this one observes that intermediate multiparticle states,
with the particles sufficiently separated, should satisfy the same selection
rules as described above by Eq. (\ref{8.3}). As a consequence, the S-matrix
elements for $N$-particle scattering amplitudes can be expressed as a
product of two-particle S-matrices.

Considering the wave packet 
\begin{equation}
\psi (x)=\int dpe^{-a(p-p_{0})^{2}+ip(x-x_{0})}|p\rangle
\end{equation}%
the action of a higher (local) charge leads to 
\begin{equation}
e^{icQ^{(n)}}\psi (x)=%
\tilde{\psi}(x)=\int dpe^{-a(p-p_{0})^{2}+ip(x-x_{0})+icp^{n}}|p\rangle ,
\label{8.6}
\end{equation}%
which is a wave packet now centered at the point $\tilde{x}_{0}$, given by $%
\tilde{x}_{0}=x_{0}-ncp_{0}^{n-1}$. The shift is proportional to a power of $%
p_{0}$, hence it grows with $p_{0}$.

It is not difficult to see that for a three particle scattering ($i,j,k$) a
momentum dependent shift implies 
\begin{equation}
S_{ij}(\theta _{ij})S_{ik}(\theta _{ik})S_{jk}(\theta _{jk})=S_{jk}(\theta
_{jk})S_{ik}(\theta _{ik})S_{ij}(\theta _{ij})\quad ,  \label{8.11}
\end{equation}%
where $\theta _{ij}$ is the rapidity 
\index{S-matrix!and rapidity variable} defined by $\theta _{ij}=\theta
_{i}-\theta _{j}$, with 
\begin{equation}
p_{i}=m(\cosh \theta _{i},\sinh \theta _{i}),  \label{8.12}
\end{equation}%
and where $m$ is the mass of the fundamental particles. A second, purely
algebraic interpretation of equation (\ref{8.11}) is also possible. We
consider the symbols $\{A_{i}(\theta )\}$ to represent the set of particles.
A given $n$-particle state is defined by the action of a product of these
symbols on the vacuum, ordered according to their rapidities: the
\textquotedblleft in" states are identified with the products in order of
decreasing rapidities, while the \textquotedblleft out" states are arranged
in the order of increasing rapidities. The commutation relations of the $%
A^{\prime }$s are defined in terms of the S-matrix, that is, 
\begin{equation}
A(\theta _{1})A^{\prime }(\theta _{2})=S_{T}(\theta _{12})A^{\prime }(\theta
_{2})A(\theta _{1})+\cdots \quad ,  \label{8.13}
\end{equation}%
where $S_{T}$ is the transition amplitude for $AA^{\prime }\rightarrow
AA^{\prime }$, and the dots represent other channels. There are different
ways to consider the scattering of three particles and uniqueness of the
result leads to equation (\ref{8.11}).

The two particle S-matrix in a factorizable two-dimensional theory is a
function of the Mandelstam variable $s$. It is convenient to write the
momenta $p_{i}$ in terms of the rapidity variable $\theta _{i}$ as defined
in (\ref{8.12}). The two-particle S-matrix elements depend on the difference
of rapidities: they only depend on the variable $s$, related to $\theta
=\theta _{i}-\theta _{j}$ by 
\begin{equation}
s=(p_{i}+p_{j})^{2}=m_{i}^{2}+m_{j}^{2}+2m_{i}m_{j}\cosh \theta \quad .
\label{8.14}
\end{equation}%
For equal masses we have $s=2m^{2}(1+\cosh \theta )=4m^{2}(\cosh {%
\frac{\theta }{2}})^{2}$.

In general the two-particle amplitudes are analytic functions of $s$, with
cuts along the real axis. The scattering amplitude has a cut for $s\leq
(m_{1}-m_{2})^{2}$, and for $s\geq (m_{1}+m_{2})^{2}$. The point $%
s=(m_{1}+m_{2})^{2}$ corresponds to the two-particle threshold. The mapping (%
\ref{8.14}) transforms the physical sheet in the $s$-plane into a strip $%
0<\Im m\theta <\pi $. The scattering amplitude $S(\theta )$ is real analytic
and hence is real on the imaginary $\theta $ axis. Moreover, on the real
axis $S(-\theta )=S^{\ast }(\theta )$.

In the calculation of S-matrices in two dimensions, one first computes the
so-called minimal S-matrix, which has a minimum number of zeros and poles on
the physical sheet and grows slower than $\exp {\frac{p.p^{\prime }}{m^2}}$
for large momenta. At this point one requires that the S-matrix obeys
unitarity and crossing \cite{[190;8]}. The first condition turns out to be a
requirement on the modulus squared of the two-particle scattering amplitude,
since there is no particle production \cite{[36;8]}.

In a relativistic theory, crossing corresponds to the substitution of an
incoming particle of momentum $p$ by an outgoing antiparticle with momentum $%
-p$. This is equivalent to the substitution $s\rightarrow 4m^{2}-s$ (or $%
\theta \rightarrow i\pi -\theta $). In terms of equations, invariance under
crossing implies 
\begin{equation}
\langle f_{1}^{\prime }f_{2}^{\prime }|S(P_{1},P_{2})|f_{1},f_{2}\rangle
=\langle f_{1}^{\prime }\overline{f}_{2}|S(P_{1},-P_{2})|f_{1}\overline{f}%
_{2}^{\prime }\rangle \quad .
\end{equation}%
Crossing symmetry leads to useful constraints on the scattering amplitudes,
and will be used frequently in order to fix the S-matrices.

We can summarize the whole program of computing exact S-matrices in the
following steps \cite{[36;8],[37;8],[39;8]}:

\begin{enumerate}
\item Set up the factorization equations, either from the local conservation
laws, such as in (\ref{8.11}), or using the non local conservation laws.

\item Impose crossing and unitarity.

\item Compute the minimal S-matrix, that is the one obeying analyticity,
having the minimum number of zeros or poles in the physical sheet, and
growing asymptotically slower than $\exp\vert {\frac{p_1p_2}{m^2}}\vert $
for $p_1,p_2 \to \infty $.

\item Using qualitative information about the bound state structure,
introduce poles; resonances are supposed to be absent, since unstable
particles do not exist for a factorizable S-matrix, due to the conservation
of the number of particles.

\item Check the results by perturbation theory, or any other method
available, as \textit{e.g.} semiclassical approximation, or $1/N$ expansion.
\end{enumerate}

As an example of factorizable S-matrix we find those with symmetry groups $%
U(N)$.

Such a symmetry requirement, implies that the particle--\-particle and
particle-antiparticle scattering amplitudes are of the form%
\begin{eqnarray}
\langle P_{\gamma }(\theta _{1}^{\prime })P_{\delta }(\theta _{2}^{\prime
})out|P_{\alpha }(\theta _{1})P_{\beta }(\theta _{2})in\rangle
&=&[u_{1}(\theta )\delta _{\alpha \gamma }\delta _{\beta \delta
}+u_{2}(\theta )\delta _{\alpha \beta }\delta _{\gamma \delta }\delta
(\theta _{1}-\theta _{2}^{\prime })\delta (\theta _{2}-\theta _{1}^{\prime })
\notag \\
&&\pm \lbrack u_{1}(\theta )\delta _{\alpha \delta }\delta _{\beta \gamma
}+u_{2}(\theta )\delta _{\alpha \beta }\delta _{\gamma \delta }\delta
(\theta _{1}-\theta _{2}^{\prime })\delta (\theta _{2}-\theta _{1}^{\prime })
\notag \\
\langle P_{\gamma }(\theta _{1}^{\prime })A_{\delta }(\theta _{2}^{\prime
})out|P_{\alpha }(\theta _{1})A_{\beta }(\theta _{2})in\rangle &=&\left[
t_{1}(\theta )\delta _{\alpha \gamma }\delta _{\beta \delta }+t_{2}(\theta
)\delta _{\alpha \beta }\delta _{\gamma \delta }\right] \delta (\theta
_{1}-\theta _{1}^{\prime })\delta (\theta _{2}-\theta _{2}^{\prime })  \notag
\\
&&\pm \left[ r_{1}(\theta )\delta _{\alpha \gamma }\delta _{\beta \delta
}+r_{2}(\theta )\delta _{\alpha \beta }\delta _{\gamma \delta }\right]
\delta (\theta _{1}-\theta _{2}^{\prime })\delta (\theta _{2}-\theta
_{1}^{\prime })  \notag
\end{eqnarray}%
where the $t$'s are transmission amplitudes and the $r$'s reflexion
amplitudes.

Implementating the factorization equations we find that the solutions fall
into six classes as given below. The function $f(\theta ,\lambda )$ is a
meromorphic function of $\theta $, for $Re\lambda >0$; it is uniquely
defined by the requirement of being minimal. The only arbitrariness lies in
the bound state structure.

\begin{itemize}
\item Class I%
\begin{eqnarray*}
r_{1}(\theta ) &=&0,\qquad t_{1}(\theta )=1,\qquad u_{1}(\theta )=1,\quad \\
r_{2}(\theta ) &=&0,\qquad t_{2}(\theta )=0,\qquad u_{2}(\theta )=0.
\end{eqnarray*}

\item Class II%
\begin{eqnarray*}
r_{1}(\theta ) &=&0,\quad t_{1}(\theta )=f(\theta ,\lambda ),\qquad
u_{1}(\theta )=t_{1}(i\pi -\theta ), \\
\quad r_{2}(\theta ) &=&0,\qquad t_{2}(\theta )=\frac{i\pi \lambda }{\theta
-i\pi }t_{1}(\theta ),\qquad u_{2}(\theta )=-\frac{i\pi \lambda }{\theta }%
u_{1}(\theta ).
\end{eqnarray*}

\item Class III%
\begin{eqnarray*}
r_{1}(\theta ) &=&-\frac{i\pi \lambda }{\theta }t_{1}(\theta ),\quad
t_{1}(\theta )=f(\theta ,\lambda )f(i\pi -\theta ,\lambda ),\qquad
u_{1}(\theta )=t_{1}(\theta ), \\
\quad r_{2}(\theta ) &=&\frac{i\pi \lambda }{\theta -i\pi }t_{1}(\theta
),\qquad t_{2}(\theta )=r_{2}(\theta ),\qquad u_{2}(\theta )=r_{1}(\theta ).
\end{eqnarray*}

\item Class IV%
\begin{eqnarray*}
r_{1}(\theta ) &=&-\frac{i\pi \lambda }{\theta }t_{1}(\theta ),\quad
t_{1}(\theta )=f(\theta ,\lambda )f(i\pi -\theta ,\lambda )i\tanh \frac{1}{2}%
(\theta +\frac{1}{2}i\pi ),\qquad u_{1}(\theta )=-t_{1}(\theta ), \\
r_{2}(\theta ) &=&\frac{i\pi \lambda }{\theta -i\pi }t_{1}(\theta ),\qquad
t_{2}(\theta )=r_{2}(\theta ),\qquad u_{2}(\theta )=r_{1}(\theta ).
\end{eqnarray*}

\item Class V%
\begin{eqnarray*}
r_{1}(\theta ) &=&{{\prod\limits_{k=-\infty }^{\infty }\!\!\!{\frac{f(\theta
,k/2\mu i)}{f(\theta ,k/2\mu i+\!{\frac{1}{2}})}}}},\qquad t_{1}(\theta
)=0,\qquad u_{1}(\theta )=0,\quad \quad \\
r_{2}(\theta ) &=&{{\frac{\sin \mu (i\pi -\theta )}{\sin \mu \theta }}%
r_{1}(\theta ),\qquad }t_{2}(\theta )=r_{2}(\theta ),\qquad u_{2}(\theta
)=r_{1}(\theta ).
\end{eqnarray*}

\item Class VI%
\begin{eqnarray*}
r_{1}(\theta ) &=&{\prod\limits_{k=-\infty }^{\infty }\!\!\!{\frac{f(\theta
,k/2\mu i)}{f(\theta ,k/2\mu i+{\frac{1}{2}})}\!\!}},\quad t_{1}(\theta
)=0,\qquad u_{1}(\theta )=0, \\
r_{2}(\theta ) &=&{{\frac{\sin \mu (i\pi -\theta )}{\sin \mu \theta }}%
r_{1}(\theta ),\qquad }t_{2}(\theta )=\mathrm{e\!\!\!\!\!\!\!}{^{i\mu (i\pi
-\theta )}r_{2}(\theta ),\qquad }u_{2}(\theta )=\mathrm{e}{^{i\mu \theta
}r_{1}(\theta ).}
\end{eqnarray*}
\end{itemize}

From these classes we see that for a $U(N)$ symmetry the solution of the
factorization equations is not unique. In the case of $\ \mathcal{C}P^{N-1}$
and chiral models, the solution will be found to be of class II; to obtain
it, we shall use the non-local conservation laws.

The solutions belonging to class III correspond to an $O(N)$ symmetry.

The chiral fermion field in two dimensions is a $SU(N)(1)\otimes \tilde{U}%
(1) $ multiplet of fermions. The Lagrangian is defined by 
\begin{equation}
\mathcal{L}=i\overline{\psi }_{i}\not{\!}\partial \psi _{i}+{\frac{1}{2}}g[(%
\overline{\psi }_{i}\psi _{i})^{2}-(\overline{\psi }_{i}\gamma _{5}\psi
_{i})^{2}],  \label{5.63}
\end{equation}%
where the summation over the $SU(N)$ index $i$ is understood. The Lagrangian
(\ref{5.63}) again defines an integrable model. The Noether current
associated with the $U(N)$ symmetry is given by 
\begin{equation}
j_{ij}^{\mu }=i\overline{\psi }_{j}\gamma ^{\mu }\psi _{i},
\end{equation}%
Use of the equation of motion and Fierz transformation, shows that it
satisfies 
\begin{equation}
\partial _{\mu }j_{\nu }-\partial _{\nu }j_{\mu }+ \frac 12 g[j_{\mu
},j_{\nu }]=0 \quad .
\end{equation}%
This relation shows the integrability of the model model, and implies the
existence of a non-local conserved charge of the usual form.

A possible candidate to exact S-matrix describing the scattering of
elementary fermions is that of class II in the Table. A strong indication of
this fact should come with the $1/N$ expansion. However, there is a massless
field in the theory if we try to obtain perturbation naively. In two
dimensional space-time this can lead to infrared divergencies very difficult
to deal with. The solution of such a problem was given by two independent
papers . We quickly review them here.

\textit{Cancellation of infrared singularities} 

In order to obtain the $1/N$ expansion of this model, we have to reformulate
it. The theory can be reduced to a quadratic form in $\psi $ at the expense
of two auxiliary fields, 
\begin{equation}
\mathcal{L}=i\overline{\psi }\not{\!}\partial \psi -{\frac{1}{2g}}(\sigma
^{2}+\pi ^{2})+\overline{\psi }(\sigma +i\pi \gamma _{5})\psi .  \label{5.67}
\end{equation}

However, the $1/N$ expansion of the model using the above Lagrangian cannot
be performed, due to serious infrared (IR) problems \cite{[196;5]}: we find
a massless pole in the $\pi $ propagator. It plays the role of the
problematic massless Goldstone boson \cite{aar}. We now rewrite the fields
in terms of $\sigma +i\pi =\rho e^{i\phi }$, leading to the Lagrangian 
\begin{equation}
\mathcal{L}=i\overline{\psi }\not{\!}\partial \psi -{\frac{1}{2g}}\rho
^{2}+\rho \overline{\psi }e^{i\phi \gamma _{5}}\psi\quad .  \label{5.68}
\end{equation}

We now discuss the quantum theory associated with the above classical
Lagrangian. The most pedestrian approach consists in the extensive use of
the bosonization formulae. The fermionic fields are bosonized in terms of an 
$N$-plet $\varphi _{i}$. The situation is analogous to the massive Thirring
model and one obtains the equivalent bosonic Lagrangian 
\begin{equation}
\mathcal{L}={\frac{1}{2}}\sum_{i=1}^{N}(\partial _{\mu }\varphi _{i})^{2}-{%
\frac{1}{2g}}\rho ^{2}+{\frac{\mu }{2\pi }}\rho \sum_{i=1}^{N}\cos (\phi
+\varphi _{i}\sqrt{4\pi }).  \label{5.69}
\end{equation}

It is now convenient to use \textquotedblleft fermionization" formulae in
order to rewrite (\ref{5.69}) in terms of new fermion fields $\tilde{\psi}%
_{i}$ \cite{[197;5],[198;5]}, by making the identifications 
\begin{eqnarray}
i\overline{\tilde{\psi}}_{i}\not{\!}\partial \tilde{\psi}_{i} &=&{\frac{1}{2}%
}\Bigl[\partial _{\mu }\Bigl(\varphi _{i}+{\frac{\phi }{\sqrt{4\pi }}}\Bigr)%
\Bigr]^{2},  \notag \\
\overline{\tilde{\psi}_{i}}\tilde{\psi}_{i} &=&{\frac{\mu }{2\pi }}\cos
(\phi +\varphi _{i}\sqrt{4\pi }),  \notag \\
\overline{\tilde{\psi}_{i}}\gamma _{\mu }\tilde{\psi}_{i} &=&-{\frac{1}{%
\sqrt{\pi }}}\epsilon _{\mu \nu }\partial ^{\nu }\left( \varphi _{i}+{\frac{%
\phi }{\sqrt{4\pi }}}\right) .
\end{eqnarray}%
The Lagrangian (\ref{5.69}) then takes the form \cite{[198;5]} 
\begin{equation}
\mathcal{L}=\overline{\tilde{\psi}_{i}}i\not{\!}\partial \tilde{\psi}_{i}-{%
\frac{1}{2g}}\rho ^{2}+\rho \overline{\tilde{\psi}_{i}}\tilde{\psi}_{i}+{%
\frac{1}{2}}\partial _{\mu }\phi \overline{\tilde{\psi}_{i}}\gamma ^{\mu
}\gamma _{5}\tilde{\psi}_{i}+{\frac{N}{8\pi }}(\partial _{\mu }\phi )^{2}.
\label{5.71}
\end{equation}

The important point is that the massless field $\phi $ interacts only via
its derivative, thus implying the absence of infra red problems in the
correlation functions of $\tilde{\psi}$. We can obtain the same Lagrangian (%
\ref{5.71}) by computing the fermionic determinant associated with the
Lagrangian (\ref{5.68}) (see \cite{aar}).


\subsection{The ${\frac{1}{N}}$ expansion}

The large $N$ expansion of the model defined by the Lagrangian (\ref{5.71})
can be explicitly performed \cite{[198;5]}. The propagator of the $\rho $
field is exactly the same as that obtained for the $\tilde \sigma $ field in
the $O(N)$ case. The zero'th order contribution to the $\rho $-propagator is
thus given by 
\begin{equation}
\tilde{\Gamma}(p)=-{\frac{i}{2\pi }}{\frac{\theta }{\tanh {\frac{\theta }{2}}%
}},  \label{5.84}
\end{equation}%
where $\theta $ is defined by $p^{2}=-4m^{2}\sinh ^{2}{\frac{\theta }{2}}$.

Since only $\partial _{\mu }\phi $ occurs in (\ref{5.71}), we just need the
two point function of $A_{\mu }=\sqrt{N}\epsilon _{\mu \nu }\partial ^{\nu
}\phi $, given by 
\begin{eqnarray}
\tilde{\Gamma}_{\mu \nu }(p) &=&{\frac{1}{2\pi }}\theta \tanh {\frac{\theta 
}{2}}(g_{\mu \nu }p^{2}-p_{\mu }p_{\nu }),
\end{eqnarray}%
where $p^{2}=-4m^{2}\sinh ^{2}{\frac{\theta }{2}}$.

The amplitudes for particle scattering are all free from IR divergencies,
and may be computed without difficulty. We can compute the two particle
scattering amplitude in lowest order \cite{aar}. The lowest order
contributions to $u_{1}(\theta )$ lead to 
\begin{equation}
u_{1}(\theta )=1+{\frac{i\pi }{N}}\coth \left( {\frac{\theta }{2}}\right) .
\label{5.89}
\end{equation}%
Moreover the backward fermion antifermion scattering vanishes, confirming
the S matrix benomging to the class II defined before.

\textit{Operator formulation}\label{sec3.3of5} 

This model may also be studied in the operator formalism, which leads to the 
$1/N$ expansion, and a correct understanding of the relation between the
``candidate" Goldstone boson and chiral symmetry.

Since the fields $\psi _{i}$ lie in the fundamental representation of $U(N)$%
, we have the bosonic representation \cite{[199;5]} 
\begin{equation}
\psi _{i}(x)=K_{i}\left( {\frac{\mu }{2\pi }}\right) ^{\frac{1}{2}}e^{-i{%
\frac{\pi }{4}}\gamma _{5}}\colon e^{i\sqrt{\frac{\pi }{N}}\left[ \gamma
_{5}\chi (x)+\int_{x^{1}}^{\infty }dy^{1}\dot{\chi}(x^{0},y^{1})\right]
}\colon \colon e^{-i\sqrt{\pi }\left[ \gamma _{5}\chi
_{i}(x)+\int_{x^{1}}^{\infty }dy^{1}\dot{\chi}_{i}(x^{0},y^{1})\right]
}\colon  \label{5.90}
\end{equation}%
with $i=1,\ldots ,N$. Since the $\chi _{i}^{\prime }s$ are $SU(N)$ valued;
they are not independent, but satisfy 
\begin{equation}
\sum_{i=1}^{N}\chi _{i}(x)=0.  \label{5.91}
\end{equation}

The field $\chi $ is the potential of the conserved $U(1)$ current. Its zero
-mass character will ensure that the $U(1)$ symmetry is not spontaneously
broken.

In the above, $K_{i}$ is a Klein factor, necessary to enforce the correct
anticommutation relations among different $\psi _{i}^{\prime }s$. Due to the 
$U(1)\times \widetilde{U}(1)$ symmetry, the divergence and the curl of the $%
U(1)$ current vanish, so that the field $\chi (x)$ is massless. Therefore
the fermion fields contain the so-called infraparticles \cite{[92;5]}, and
we need to extract them in order to arrive at the physical fields of the
theory. They are given by 
\begin{equation}
\hat{\psi}_{i}(x)=K_{i}\sqrt{{\frac{\mu }{2\pi }}}e^{i\sqrt{\pi }\left\{
\gamma _{5}\chi _{i}(x)+\int_{x^{1}}^{\infty }dy^{1}\dot{\chi}%
_{i}(x^{0},y^{1})\right\} }.  \label{5.92}
\end{equation}

The $\psi $ fields (\ref{5.92}) will be found to correspond to the field $%
\widetilde{\psi }_{i}$ in (\ref{5.71}). These fields no longer carry $%
U(1)\times \widetilde{U}(1)$ charge, and transform as a representation of $%
SU(N)$. The constraint (\ref{5.91}) implies 
\begin{equation}
{\hat{\psi}^{\dagger }}_{i}\sim {\frac{1}{(n-1)!}}\epsilon _{ii_{1}\cdots
i_{n-1}}\hat{\psi}_{i_{1}}\cdots \hat{\psi}_{i_{N-1}},  \label{5.93}
\end{equation}%
where on the right hand side a suitable redefinition of the Klein factor and
the normal product prescription is required. Eq. (\ref{5.93}) states that
the antifermions of the chiral Gross--Neveu model can be viewed as a bound
state of $N-1$ fermions. We use this fact to determine the S-matrix and its
pole structure.

Asymptotically, one expects $\hat{\psi}$ to describe massive particles, so
that one should have \cite{[199;5]} 
\begin{equation}
\hat{\psi}(vt,t)\rightarrow {\frac{1}{\sqrt{|t|}}}\{e^{-im\gamma ^{-1}t}\hat{%
a}(m\gamma v)+e^{im\gamma ^{-1}t}\hat{b}^{\dagger }(m\gamma v)\},
\label{5.94}
\end{equation}%
where $\gamma ={\frac{1}{\sqrt{1-v^{2}}}}$.

The fields $\hat{\psi}_{i}$ carry spin $s={\frac{1}{2}}(1-1/N)$, 
\begin{equation}
\hat{\psi}(x,t)\hat{\psi}(y,t)=e^{2\pi is\epsilon (x-y)}\hat{\psi}(y,t)\hat{%
\psi}(x,t),
\end{equation}%
implying an unusual \textsl{statistics} for the creation and annihilation
operators defined in (\ref{5.94}) 
\begin{equation}
\hat{a}^{\dagger }(p)\hat{a}^{\dagger }(p^{\prime })=e^{2\pi is\epsilon
(p-p^{\prime })}\hat{a}^{\dagger }(p^{\prime })\hat{a}^{\dagger }(p).
\end{equation}

Since no scattering theory is known for particles with the above statistics,
it is necessary to replace the field $\hat{\psi}$ by another field $\psi
^{\prime }$ with a well defined statistics. This is achieved by introducing
in (\ref{5.90}) free massless scalar and pseudoscalar fields $B$ and $A$,
quantized with metric opposite to that of $\chi (x)$, in such a way that the
divergent infrared behavior of $\psi $ induced by $\chi (x)$ is compensated,
without affecting the statistics. We define \cite{[199;5]} 
\begin{equation}
\psi _{i}^{\prime }(x)=e^{i\sqrt{{\frac{\pi }{N}}}[\gamma
^{5}A(x)+B(x)]}\psi _{i}(x).  \label{5.97}
\end{equation}%
Correspondingly, the 
operators ${a}^{\dagger },a,{b}^{\dagger },b$ are related to $\hat{a}%
^{\dagger },\hat{a},\hat{b}^{\dagger },\hat{b}$ by 
\begin{eqnarray}
\hat{a}_{in}^{\dagger }(p) &=&a_{in}^{\dagger }(p)e^{2\pi i\left( s-{\frac{1%
}{2}}\right) \int_{p}^{\infty }N_{in}(p^{\prime })dp^{\prime }},  \notag \\
\hat{a}_{out}^{\dagger }(p) &=&a_{out}^{\dagger }(p)e^{2\pi i\left( s-{\frac{%
1}{2}}\right) \int_{\infty }^{p}N_{out}(p^{\prime })dp^{\prime }}.
\end{eqnarray}%
where $N_{\atop{}{{in}{out}}}$ are the corresponding particle number
operators.

Since we expect $\psi _{i}^{\prime }(x)$ to be a local field describing
massive degrees of freedom, we should have in the far past and future \cite%
{[204;5]} 
\begin{equation}
\psi ^{\prime }(vt,t)\rightarrow {\frac{1}{\sqrt{|t|}}}[e^{-im\gamma
^{-1}t}a_{\atop{}{{}{{out}{in}}}}(m\gamma v)+e^{im\gamma ^{-1}t}b_{\atop{}{%
{}{{out}{in}}}}^{\dagger }(m\gamma v)].
\end{equation}%
Substitution of $\psi $ in terms of $\psi ^{\prime }$ in (\ref{5.63}) leads
formally to the Lagrangian 
\begin{eqnarray}
\mathcal{L} &=&i\overline{\psi }_{i}^{\prime }\not{\!}\partial \psi
_{i}^{\prime }+{\frac{1}{2}}g[(\overline{\psi }_{i}^{\prime }\psi
_{i}^{\prime })^{2}-(\overline{\psi }_{i}^{\prime }\gamma _{5}\psi
_{i}^{\prime })^{2}]-{\frac{1}{2}}(\partial _{\mu }A)^{2}  \notag \\
&&-{\frac{1}{2}}(\partial _{\mu }B)^{2}+{\frac{\alpha }{\sqrt{N}}}\overline{%
\psi }^{\prime }\gamma ^{5}\gamma ^{\mu }\psi ^{\prime }\partial _{\mu }A-{%
\frac{\beta }{\sqrt{N}}}\overline{\psi }^{\prime }\gamma ^{\mu }\psi
^{\prime }\partial _{\mu }B,  \notag \\
&&\quad  \label{5.101}
\end{eqnarray}%
where we allowed for general couplings $\alpha $ and $\beta $, which after
renormalization should reduce to $\sqrt{\pi }$ as the renormalized value. We
will came back to this point after obtaining the $\frac{1}{N}$ expansion,
which we consider next.

The effective action obtained from the Lagrangian (\ref{5.101}) after
introduction of the auxiliary fields $\sigma $ and $\pi $ (compare with (\ref%
{5.67})) is given by 
\begin{eqnarray}
S_{eff} &=&-iN\mathrm{tr}\ln \left\{ i\not{\!}\partial +\sigma +i\pi \gamma
_{5}+{\frac{\alpha }{\sqrt{N}}}\gamma ^{5}\not{\!}\partial A-{\frac{\beta }{%
\sqrt{\pi }}}\not{\!}\partial B\right\}  \notag \\
&-&{\frac{1}{2g}}\int d^{2}x(\sigma ^{2}+\pi ^{2})-{\frac{1}{2}}\int
d^{2}x[(\partial _{\mu }A)^{2}+(\partial _{\mu }B)^{2}].  \notag \\
&&
\end{eqnarray}%
The field $\sigma $ is found to have a non-vanishing vacuum expectation
value $\langle \sigma \rangle =-m$, so that it is convenient to write 
\begin{equation}
\sigma =-m+{\frac{\tilde{\sigma}}{\sqrt{N}}},\quad \mathrm{and}\quad \pi ={%
\frac{\tilde{\pi}}{\sqrt{N}}}.
\end{equation}%
The second order contribution to the effective action can be computed and
the $1/N$ expansion turns out to be well defined.

We fix the parameters $\alpha $ and $\beta $ in (\ref{5.101}) by requiring
that the IR divergencies cancel. We expect to obtain for the
non-renormalized values, $\alpha =\infty $ (corresponding to $\alpha _{ren}=%
\sqrt{\pi }$ and $\beta =\beta _{ren}=\sqrt{\pi }$, since $B$ couples to a
conserved current).

We have thus verified in the $N\to \infty $ limit that both Lagrangians (\ref%
{5.71}) and (\ref{5.101}) lead to the same result. In the limit $\alpha \to
\infty $, the renormalized coupling $\alpha _{ren}$ indeed turns out to be $%
\sqrt \pi $, as one reads off from the four-point function, and the pole in
the $\pi $-propagator vanishes. To summarize, we conclude that part of the
field (\ref{5.90}) which carries chirality decouples \cite%
{[197;5],[198;5],[199;5]} from the physical spectrum and the remaining part
describes an $SU(N)$ multiplet with a well defined factorizable S-matrix.


\subsection{Quantization of non-local charge}

The discussion of the existence and conservation of a non-local charge in
the quantum chiral Gross--Neveu model follows exactly the same pattern as in
the $O(N)$ invariant model. No anomaly exists in this case. It is not
difficult to see that the action of the charges on asymptotic states is
given in this case by {\footnotesize 
\begin{eqnarray}
Q^{ab}|\theta _{1}i;\theta _{2}j\rangle &=&|\theta _{1}k;\theta _{2}l\rangle %
\bigl[-(I^{ac})^{ik}(I^{cb})^{jl}+{\frac{N}{i\pi }}\theta
_{1}(I^{ab})^{ik}\delta ^{jl}+{\frac{N}{i\pi }}\theta
_{2}(I^{ab})^{jl}\delta ^{ik}\bigr],  \notag \\
\langle \theta _{1}i;\theta _{2}j|Q^{ab}&=&\langle \theta _{1}k;\theta _{2}l|%
\bigl[-(I^{ac})^{ki}(I^{cb})^{lj}+{\frac{N}{i\pi }}\theta
_{1}(I^{ab})^{ki}\delta ^{lj}+{\frac{N}{i\pi }}\theta
_{2}(I^{ab})^{lj}\delta ^{ik}\bigr],  \notag \\
Q^{ab}|\theta _{1}i;\overline{\theta }_{2}j\rangle &=&|\theta _{1}k;%
\overline{\theta }_{2}l\rangle \bigl[-(I^{ac})^{ik}(I^{cb})^{jl}+{\frac{N}{%
i\pi }}\theta _{1}(I^{ab})^{ik}\delta ^{jl}-{\frac{N}{i\pi }}\overline{%
\theta }_{2}(I^{ab})^{lj}\delta ^{ik}\bigr],  \notag \\
\langle \theta _{1}i;\overline{\theta }_{2}j|Q^{ab}&=&\langle \theta _{1}k;%
\overline{\theta }_{2}l|\bigl[-(I^{ac})^{ki}(I^{cb})^{lj}+{\frac{N}{i\pi }}%
\theta _{1}(I^{ab})^{ki}\delta ^{lj}-{\frac{N}{i\pi }}\overline{\theta }%
_{2}(I^{ab})^{jl}\delta ^{ik}\bigr],  \notag
\end{eqnarray}%
} where $I^{ab}$ are the $SU(N)$ generators\cite{aar}. Conservation of the
charge leads to the factorization equations and to the exact S matrix of the
problem. 
\textit{First Conclusions and Physical Interpretation} 

The Gross--Neveu models are simple but physically rich models. The
semi-classical analysis, both in the $O(N)$ and in the $SU(N)\times
U(1)\times \tilde{U}(1)$-symmetric cases reveals that the models have a rich
bound-state structure \cite{[180;5]}.

The chiral Gross--Neveu model is particularly interesting, due to the chiral
symmetry breaking issue. Since, as we saw in the previous section, a mass
term is dynamically generated for the fermion, one could be led to conclude
that the chiral symmetry is broken, which is prohibited in two-dimensional
space-time. This problem has been discussed at length by several authors 
\cite{[197;5],[198;5]}. The interesting outcome is that the chirality
carrying field decouples from the theory (Eq. (\ref{5.97})). In the operator
language, this is realized by the factorization of the auxiliary fields $A$
and $B$. The physical fermions, as given by either (\ref{5.92}) or (\ref%
{5.97}), though exhibiting a non-vanishing mass gap, are chiral singlets.
This physical picture is carried over to the supersymmetric $\mathcal{C}%
P^{N-1}$ model, and reflects the fact that antiparticles are bound states of
particles, in both, the Gross--Neveu model, and in the supersymmetric $%
\mathcal{C}P^{N-1}$ model. This permits the computation of the S-matrix for
these two cases \cite{[206;5]}.

\section{The Exact solutions of classes of Integrable Models 
and String Theories}


Large $N$ Yang-Mills theory has been frequently studied since the first
seminal paper by 't Hooft \cite{thooft}. Some time ago, it has been
discovered that there is a large $N$ limit in $\mathcal{N} =4$
supersymmetric Yang-Mills which corresponds to type IIB string theory. More
recently, we learnt from \cite{bmn} how to get the spectrum from the gauge
theory counterpart. The fact that the spectrum is related to the hamiltonian
of an integrable model \cite{beisert} is an outstanding achievement.

The integrable model is obtained from the matrix describing the anomalous
dimensions of certain classes of fields in super Yang Mills theory, in the
field theory counterpart.

The procedure is obtained from the renormalization group equation 
\begin{equation}
\left\lbrace\mu\frac{\partial}{\partial\mu}+ \gamma\right\rbrace \Gamma = 0
\end{equation}
where $\Gamma$ describes the correlator of the fields under study and $\mu$
is a renormalization group parameter. As it turns out, $\gamma$ describes a
matrix valued Hamiltonian whose indices describe the different fields in the
correlator, and its diagonalization amounts to a solution of an integrable
model.

Such a statement is a very nontrivial fact about some field theories
relating them in a very remarkable fashion. Indeed, the existence of
integrable structures in gauge theories, at classical as well as quantum
level, in two and four dimensional space-time has been suspected long ago in
different setups \cite{75,several} and a huge amount of more recent
literature concerning integrable structures in string related theories have
appeared \cite{bmn,dewolfe,others}.

Here we discuss boundaries in open spin chains with $SO(6)$ symmetry and
their corresponding interpretation in super Yang-Mills theory with four
supercharges. Furthermore the spin chain with static boundary conditions has
a more general parameter space, which may suggest a larger class of
operators whose one loop anomalous dimension matrix is correspond to an
integrable spin chain. Here the most general $SO(6)$ open spin chain
Hamiltonian will be proposed using integrability requirements.

The $AdS/CFT$ (Anti-de Sitter/ Conformal Field Theory) conjecture relates
two very different theories in two very different settings, this is why at
fist the conjecture seems so surprising and interesting. In one side of the
conjecture we have a quantum theory of gravity in an assymptoticaly AdS
space and the other we a conformal quantum field theory in the boundary of
the AdS space, which is the standard Minkowski space. The claim is that for
every observable in one side of the conjecture there is a corresponding
observable in the other side of it. Gauge invariant single trace operators
in the quantum field theory side corresponds to physical states in the
quantum gravity side. And correlation functions (there is no S-matrix in a
CFT) in the quantum field theory are calculated using quantum gravity states
with appropriate boundary conditions.

The possible objects to compare in both sides are not limited to states and
correlation functions. There is a very large amount of evidence for this
conjecture and we refer to \cite{adsreview} for the most important ones. The
best known example is the case of Type IIB string theory in $AdS_5\times S^5$
space which is dual to $N=4$ SYM theory in four dimensions\cite{ads5s5}.
This case is particularly interesting since it preserves all possible
supersymmetries in ten dimensions and has the largest possible symmetry
algebra in four dimensions.

The issue that prevents a better understanding of this conjecture is that
the sigma models describing the dynamics of the string in such backgrounds
is a complicated CFT. Although these sigma models appear to be integrable 
\cite{integrable} (they have an infinite number of conserved charges), no
one was able to use the integrable structure to make any non-trivial
computation. There are many questions regarding this problem, for example,
integrable field theories in d=2 usually have a mass gap, but in the case at
hand there is no S-matrix. At least in the first order of perturbation
theory it was shown that there is no particle production, a property of
integrable field theories. On the other hand, there was much progress in the
super Yang-Mills side of the conjecture.

\subsection{N=4 supersymmetric Yang-Mills Theory}

In four dimensions there is only one field theory with 16 supercharges that
does not contain gravity: N=4 supersymmetric Yang-Mills theory with coupling
constant $g$ and gauge group $SU(N)$ Other gauge groups are allowed, but
will not be considered here. This theory is unique up to the choice of the
gauge group and coupling constant. Its field content is the gauge field $%
A^{\mu }$, 4 fermions in the fundamental representation of $SU(4)$ (the
R-symmetry group) $\psi _{\alpha }^{A}$, where $A$ is an $SU(4)$ index and $%
\alpha $ is a spinor index and there are 6 scalars in the antisymmetric
representation of $SU(4)$ $\phi ^{AB}$. The scalars can also be seen as
vectors of $SO(6)$ and the fermions as spinors of the same group. We can use
the gamma matrices $\gamma _{AB}^{I}$ to transform one representation into
the other.

The lagrangean of this theory ignoring terms with fermions is 
\begin{eqnarray}
S &=&\int d^{4}x\mathrm{Tr}[{\frac{1}{4}}F^{\mu \nu }F_{\mu \nu }+{\frac{1}{2%
}}D_{\mu }\phi ^{AB}D^{\mu }\phi _{AB}  \notag \\
&&+{\frac{1}{4}}g^{2}[\phi ^{I},\phi ^{J}][\phi _{I},\phi _{J}]+\cdots
\label{symaction}
\end{eqnarray}%
The underlying symmetry group of this theory is very large. The classical
conformal invariance is not broken in the quantum theory. The conformal
transformations together with the super Poincar\'{e} group form the algebra $%
PSU(2,2|4)$, with 30 bosonic (including the R-symmetry generators) and 32
fermionic generators. Among all this symmetries, a especial one is the scale
symmetry, generated by the dilatation operator $\mathcal{D}$, to be defined
later on.

\subsection{Single Trace Operators}

One class of interesting observables in this theory are the gauge invariant
single trace operators. The most obvious example is 
\begin{equation}
\mathcal{O}_{F}=\mathrm{Tr}(F^{\mu \nu }F_{\mu \nu }).
\end{equation}%
In the $AdS/CFT$ correspondence this operator couples to the dilaton.
Therefore it corresponds to a change in the coupling constant. This is an
example of a chiral operator as well, since it is annihilated by half of the
supercharges (half of the supersymmetry generators \textit{and} half of the
superconformal transformations). This operator is a descendant of 
\begin{equation}
\mathcal{O}_{\phi }=\mathrm{Tr}(\phi ^{\{I}\phi ^{J\}}),
\end{equation}%
where $\{IJ\}$ means symmetric traceless combination. This means it can be
obtained from the above expression by means of the action of some
supercharges. All chiral single trace operators with only two fields can be
obtained from the one above from the action of the supercharges. More
generally, \textit{all} chiral operators operators are obtained from 
\begin{equation}
\mathcal{O}_{n}=\mathrm{Tr}(\phi ^{\{I_{1}}\phi ^{I_{2}}\cdots \phi
^{I_{n-1}}\phi ^{I_{n}\}}).
\end{equation}%
In summary, all chiral operators in $N=4$ super Yang-Mills are related to
massless states in the corresponding string theory. The spectrum of these
operators is easy to compute, since they are protected by quantum
corrections. The dimensions are the classical ones, which can be easily
computed from the classical action. A much more difficult problem, which has
not yet been completely solved is the computation of the dimension of 
\textit{any} gauge invariant single trace operator. The most important
progress on this problem is the conjecture that the dimension of any gauge
invariant operator is an eigenvalue of the Hamiltonian of some integrable
spin chain. The simplest example of a non chiral operator is the Konishi
operator 
\begin{equation}
\mathcal{O}_{K}=\mathrm{Tr}(\phi ^{I}\phi ^{I}).
\end{equation}%
Its one loop anomalous dimension can be computed using standard methods and
does not vanish.

\subsection{Dilatation Operator and Spin Chain Hamiltonian}

In field theory the dilatation operator $\mathcal{D}$ gives the conformal
dimension (classical plus anomalous dimension) upon commutation, by means of
the expression 
\begin{equation}
\lbrack \mathcal{D},\mathcal{O}]=\Delta _{\mathcal{O}}\mathcal{O},
\end{equation}%
whenever we have a diagonal base, $\Delta $ being the conformal dimension.
The more general situation is 
\begin{equation}
\lbrack \mathcal{D},\mathcal{O}_{i}]=\Delta _{ij}\mathcal{O}_{j},
\end{equation}%
where $\Delta _{ij}$ is the matrix of anomalous dimensions.

In a conformal field theory (CFT) this knowledge allows one to compute any
two point function, since the latter is fixed, in the simple case of scalar
operators, to be 
\begin{equation}
\langle \mathcal{O}_{i}(x)\mathcal{O}_{j}(y)\rangle ={\frac{\delta _{ij}}{%
|x-y|^{2\Delta _{i}}}}.
\end{equation}%
Thus, knowing the conformal dimensions is a small step towards a solution of
the full quantum field theory. Three point functions can also be obtained,
but more knowledge is necessary.

\subsection{The $SO(6)$ Spin chain}

The problem of studding the full $PSU(2,2|4)$ spin chain is too broad for
our proposes here. Thus, we shall review the results of the one-loop
anomalous dimension and the spin chain for the $SO(6)$ sector, which is
closed at one-loop. We refer to \cite{zarembo}.

Our interest relies in operators of the form 
\begin{equation}
\mathcal{O}_{n}=\sigma _{I_{1}I_{2}\cdots I_{n}}\mathrm{Tr}(\phi
^{I_{1}}\phi ^{I_{2}}\cdots \phi ^{I_{n}}),  \label{allops}
\end{equation}%
where $\sigma _{I_{1}I_{2}\cdots I_{n}}$ are constant polarizations. At one
loop level these operators do not mix with other types, and we can use only
the first line of Eq.\ref{symaction} to perform computations. Thus,
supersymmetry is not directly responsible for integrability at least at one
loop level. Note that we are not imposing any condition on the above
operator.

Although $N=4$ super Yang-Mills is a finite theory, some renormalization has
to be done. We only need a wave function renormalization, what is
responsible for the change of the classical dimension. We define 
\begin{equation}
\Gamma _{\mathcal{O}}=\Lambda {\frac{{\partial Z_{\mathcal{O}}}}{\partial
\Lambda },}\quad
\end{equation}%
in the simplest case. The task of computing $\Gamma _{\mathcal{O}}$ using (%
\ref{symaction}) for the operators (\ref{allops}) at one loop level has been
explained in \cite{zarembo}. The matrix of anomalous dimensions is given by 
\begin{equation}
\Gamma _{\mathcal{O}}=\lambda \sum_{l=1}^{L}(K_{l,l+1}-2P_{l,l+1}+2),
\end{equation}%
where $K_{l,l+1}$ is the trace operator and $P_{l,l+1}$ is the permutation
operator. This matrix was identified with the integrable Hamiltonian of an $%
SO(6)$ spin chain.

Using the Bethe Ansatz\cite{rational} (see also \cite{lima}) to find the
eigenvalues of this Hamiltonian one finds 
\begin{equation}
\gamma _{\mathcal{O}}=\lambda \sum_{i=1}^{n}\frac{1}{x_{i}+1/4},
\end{equation}%
where $n$ is the number of particle-like excitations and $x_{i}$ are the
rapidity parameters. We shall see that the addition of boundaries does not
change these eigenvalues, although it will put restriction on possible
operators and will change the Bethe equations.

\subsection{Solutions with Boundaries}

We now discuss how boundaries may appear in the spin chain and in the gauge
invariant operators. The single trace operators in the Yang-Mills theory are
dual to closed string states. Open strings will appear in the conjecture if
there are D-branes in the theory. The D-Brane states, or giant gravitons,
are represented by determinant operators 
\begin{equation}
\mathcal{O}_{GG}=\mathrm{det}(Z),
\end{equation}%
where $Z=\phi _{1}+i\phi _{6}$ is a highest weight state in $SO(6)$
representation and the determinant is in the adjoint representation of $%
SU(N) $. We shall attach an open string to such state. we remove one $Z$ in
the determinant above and replace it with a string of operators. To be more
explicit, the determinant is of the form 
\begin{equation}
\mathcal{O}_{GG}=\epsilon ^{j_{1}\cdots j_{N}}\epsilon _{i_{1}\cdots
i_{N}}Z_{j_{1}}^{i_{1}}\cdots Z^{i_{N}\cdots j_{N}}\quad ,
\end{equation}%
and we attach the \textquotedblleft open string\textquotedblright\ $(\psi
_{1}\cdots \psi _{L})_{j}^{i}$ to the giant graviton as 
\begin{equation}
\mathcal{O}_{o}=\epsilon ^{j_{1}\cdots j_{N}}\epsilon _{i_{1}\cdots
i_{N}}Z_{j_{1}}^{i_{1}}\cdots Z^{i_{N-1}\cdots j_{N-1}}(\psi _{1}\cdots \psi
_{L})_{j_{N}}^{i_{N}}\quad ,
\end{equation}%
where $\psi _{a}$ is one the other scalar fields. Berenstein and V\'{a}zquez
have shown that the anomalous dimension matrix for operators of this type
corresponds to the Hamiltonian of an open spin chain with static boundary
conditions. They analysed the behaviour of wave functions of this
Hamiltonian, and it was shown that the boundary conditions for elementary
excitations satisfy Dirichlet boundary conditions.


In this section the most general $SO(6)$-invariant spin chain with open
static boundary conditions will be derived.

We start with some definitions. The $SO(6)$ invariant rational $R$ matrix is
given by \cite{rational} 
\begin{equation}
R(\theta )=\left( 1-\frac{\theta }{2}\right) I+\theta \left( \frac{\theta }{2%
}-1\right) P+\frac{\theta }{2}K,  \label{R}
\end{equation}%
which satisfy the permutated Yang-Baxter equation 
\begin{equation}
R_{12}(\theta )R_{23}(\theta +\theta ^{\prime })R_{12}(\theta ^{\prime
})=R_{23}(\theta ^{\prime })R_{12}(\theta +\theta ^{\prime })R_{23}(\theta ).
\label{YBE2}
\end{equation}

These operators are explicitly represented by 
\begin{equation}
I=\sum_{i,j=1}^{6}\hat{e}_{ii}\otimes \hat{e}_{jj},\ P=\sum_{i,j=1}^{6}\hat{e%
}_{ij}\otimes \hat{e}_{ji},\ K=\sum_{i,j=1}^{6}\hat{e}_{i^{\prime }j}\otimes 
\hat{e}_{ij^{\prime }},
\end{equation}%
where $i^{\prime }=7-i$ and $(\hat{e}_{ij})_{\alpha \beta }=\delta _{i\alpha
}\delta _{j\beta }$ are standad $6\times 6$ Weyl matrices.

Using the S-matrix language we can define $S(\theta )=PR(\theta )$, in order
to recover the Yang-Baxter equation (\ref{8.11}) from the\ R-matrix equation
(\ref{YBE2}).

Following Sklyanin\cite{SK}, it turns out that an integrable $SO(6)$ open
spin chain can be obtained from the double-row transfer matrix defined as
the following trace over the $6\times 6$ auxiliary space $\mathcal{A}$,%
\begin{equation}
T(\theta )=tr(K_{{\mathcal{A}}}^{+}(\theta )\hat{R}_{{\mathcal{A}}L}(\theta
)...\hat{R}_{{\mathcal{A}}1}(\theta )K_{{\mathcal{A}}}^{-}(\theta )\hat{R}_{{%
\mathcal{A}}1}(\theta )...\hat{R}_{{\mathcal{A}}L}(\theta )).  \label{drtm}
\end{equation}

While the operator $R_{\mathcal{A}j}(u)$ determines the dynamics of the
bulk, the $6\times 6$ matrices $K_{\mathcal{A}}^{\pm }(u)$ describe the
interactions at the ends of the open chain. Moreover, compatibility with the
bulk integrability demands these matrices to satisfy the reflection
equation, which for $K_{\mathcal{A}}^{-}(u)$ reads 
\begin{equation}
R_{12}(\theta -\mu )K_{1}^{-}(\theta )R_{12}(\theta +\mu )K_{1}^{-}(\mu
)=K_{1}^{-}(\mu )R_{12}(\theta +\mu )K_{1}^{-}(\theta )R_{12}(\theta -\mu ).
\label{bybe}
\end{equation}%
while a dual equation should also hold for the matrix $K^{+}(u)$. Here $%
K_{1}^{-}(u)=K^{-}(u)\otimes I$ and $K_{2}^{-}(u)=I\otimes K^{-}(u).$

The solutions of the reflection equation (\ref{bybe}) for the $SO(6)$ $R$%
-matrix (\ref{R}) were derived in \cite{Lima}. Here we will consider only
the particular solution, 
\begin{equation}
K^{-}(\theta )=\mbox{diag}(k_{11}^{-}(\theta ),...,k_{66}^{-}(\theta ))
\label{k1}
\end{equation}%
where 
\begin{eqnarray}
k_{11}^{-}(\theta ) &=&1  \notag \\
k_{22}^{-}(\theta ) &=&\cdots =k_{55}^{-}(\theta )=-\frac{p_{-}\theta -1}{%
p_{-}\theta +1}\;\;  \notag \\
k_{66}^{-}(\theta ) &=&\frac{p_{-}\theta -1}{p_{-}\theta +1}\ \frac{%
p_{-}(\theta +1)-1}{p_{-}(\theta -1)+1}  \label{k1a}
\end{eqnarray}%
where $p_{-}$ is a free parameter.

The diagonal matrix $K^{+}(\theta )$ is obtained from crossing symmetry $%
\theta \rightarrow -\theta +2$. It turns out that the matrix elements of $%
K^{+}(\theta )$ are given by 
\begin{eqnarray}
k_{11}^{+}(\theta ) &=&1  \label{kp1} \\
k_{22}^{+}(\theta ) &=&\cdots =k_{55}^{+}(\theta )=-\frac{p_{+}(-\theta +2)-1%
}{p_{+}(-\theta +2)+1} \\
k_{66}^{+}(\theta ) &=&\frac{p_{+}(-\theta +2)-1}{p_{+}(-\theta +2)+1}\ 
\frac{p_{+}(-\theta +3)-1}{p_{+}(-\theta -1)+1}
\end{eqnarray}%
in (\ref{k1a}) and $p_{+}$ is a second free parameter.

Associated to the double-row transfer matrix (\ref{drtm}) we find the
following open spin chain Hamiltonian which is proportional to the first
order expansion of $T(\theta )$ in the spectral parameter\cite{SK}. 
\begin{eqnarray}
\mathcal{H} &=&-\sum_{i=1}^{L-1}P_{i,i+1}+\frac{1}{2}%
\sum_{i=1}^{L-1}E_{i,i+1}+\frac{1}{2}\frac{d\left( K^{-}(\theta )\right) }{%
d\theta }|_{\theta =0}+\frac{\mathrm{tr}\left( K^{+}(0)H_{L,0}\right) }{%
\mathrm{tr}\left( K^{+}(0)\right) }  \notag \\
&&  \label{ham}
\end{eqnarray}%
where $H_{i,i+1}=-P_{i,i+1}+\frac{1}{2}E_{i,i+1}$.

In order to obtain the spectrum of (\ref{ham}) in a non-perturbative way we
proceed with the exact diagonalization of the double-row operator (\ref{drtm}%
). Since the $K$-matrices considered here are diagonal, this problem can be
tackled by means of the boundary algebraic Bethe ansatz in the lines of \cite%
{GUA}.

\section{Conclusions}

There is a vast literature about the relation between four dimensional gauge
theories and two dimensional integrable models. First arised the relation
between Yang-Mills theory and two dimensional sigma models, latter some
papers appeared implying an important relation of four dimensional QCD at
high energies and spin systems in two dimensions, and a third group, more
recently, about the relation of large number of colours QCD, string theories
and integrable models, which is the basic concern of the present paper \cite%
{several}. While the first group of relations points into general
coincidences and paralels between the two classes of models, the latter two
classes of relations are definite identifications of four dimensional
physical operators and correlators with their two dimensional counterparts.

In the second case above, the large $N$ scattering in four dimensional QCD
at high energies in the leading logarithm approximation (LLA) is described
by a nearest neighbour hamiltonian equivalent to that of the Heisenberg spin
chain. Such properties have been discovered in the framework of a Feynman
diagrammatic expansion \cite{chengwu}. Later it has been argued that (3+1)
dimensional coordinates can be split into fast (with large Fourier
transform) and slow variables, and Lorentz contraction in the direction of
the motion of the fast particles rendered the corresponding field strength
to the form of a shock wave nonvanishing only in the direction of a
hyperplane passing through the trajectory of the particle.

Here the problem is even more sophisticated, relying on further properties
of the string/field theory duality. The field theory correlators of some
operators have anomalous dimension matrices corresponding to integrable
model Hamiltonians. The latter have not only familiar structures, but also
display further interesting properties concerning deformation and especially
perturbations by boundary operators. Such boundary operators can be
understood, in the string theory counterpart, as perturbing branes. In our
problem these are actually zero branes, namely, point particle operators
which do not break the original symmetries of the problem. 

We also presented the most general $SO(6)$ spin chain with open static
boundary conditions. We expected that this type of spin chain can be
associated with the one loop anomalous dimension matrix of giant graviton
operators in SYM theory\cite{berevaz}. The Hamiltonian found in the present
paper is more general than the one found previously in the literature in the
sense that it has more general boundary conditions. It would be interesting
to have an interpretation of these boundary conditions in terms of giant
graviton and D-branes in the $AdS/CFT$ duality.


We have established a perspective relating work performed in the seventies
and eighties to modern developments in string theory. The fact that today
several pieces of information from the dynamical knowledge of two
dimensional field theory is used in the search of structure in string and
superstring theories as well as super Yang Mills model shows that the models
discussed in this paper are not only relevant from the point of view of a
theoretical laboratory, but as standard tools in the search for realistic
field theories. That is the case of integrable models in the structure of
Yang Mills fields as well as string theory. The Bethe Ansatz solutions are
used to obtain the structure of anomalous dimensions and further integrable
structures in two dimensional can also be used in order to achieve knowledge
about the structure of analogous structures in the very important $%
AdS\otimes S_{5}$ space in string theory. We are also sure that much of the
dynamical structure of two dimensional models, such as that discussed in the
framework of the chiral fermion model has not been fully used as an
interesting full fledged dynamical model.

\begin{acknowledgments}
ALS wishes to thank Dr. W. Galleas for useful discussions. This work has
been supported by FAPESP and CNPQ, Brazil.
\end{acknowledgments}


\end{document}